\documentclass[12pt]{article}
\usepackage{epsfig}

\def\beq{\begin{equation}}
\def\eeq#1{\label{#1}\end{equation}}
\def\eeqn{\end{equation}}

\def\beqa{\begin{eqnarray}}
\def\eeqa#1{\label{#1}\end{eqnarray}}
\def\eeqan{\end{eqnarray}}

\let\bar=\overbar

\def\Dslash{\not{\hbox{\kern-4pt $D$}}}
\def\dslash{\not{\hbox{\kern-2pt $\del$}}}

\def\msb{{\bar{\ssstyle M \kern -1pt S}}}

\def\Title#1{\begin{center} {\Large {\bf #1} } \end{center}}

\begin{document}

\Title{Electroweak constraints from HERA}

\bigskip\bigskip


\begin{raggedright}  
{\it Elisabetta Gallo\index{Gallo, E.}\\
INFN Firenze\\
Via G. Sansone 1\\
I-50019 Sesto Fiorentino, ITALY}
\bigskip\bigskip
\end{raggedright}

\section{Introduction}

The $ep$ accelerator HERA at the laboratory DESY in Hamburg terminated 
activity on 30th June 2007. The two experiments H1 and ZEUS collected each
$0.5~\mathrm{fb}^{-1}$ of deep inelastic scattering (DIS) $ep$ collisions  at 
$\sqrt{s}=300,319~\mathrm{GeV}$, for a total of  $1~\mathrm{fb}^{-1}$ 
of integrated luminosity for
combined results. 
The data taking can be divided in two periods: the HERA I period,
corresponding to the years 1994-2000, in which each experiment collected mainly $e^+p$
data  ($\simeq 110~\mathrm{pb}^{-1}$ for $e^+p$ and 
$\simeq 15~\mathrm{pb}^{-1}$ for $e^-p$); the HERA II period, corresponding
to the years 2004-2007, after the machine performed a luminosity upgrade 
and installed spin rotators to provide longitudinally polarised lepton beams.
Approximately equal amounts of integrated luminosity for electron- and positron-proton collisions, 
with either left- or right-handed
polarisation, were provided by the end of data taking. 
The detectors are now dismantled but there is intense
activity for the publication of results on the whole statistics, as most
of the analyses published up to now involve only the HERA I data.

\begin{figure}[htb]
\begin{center}
\epsfig{file=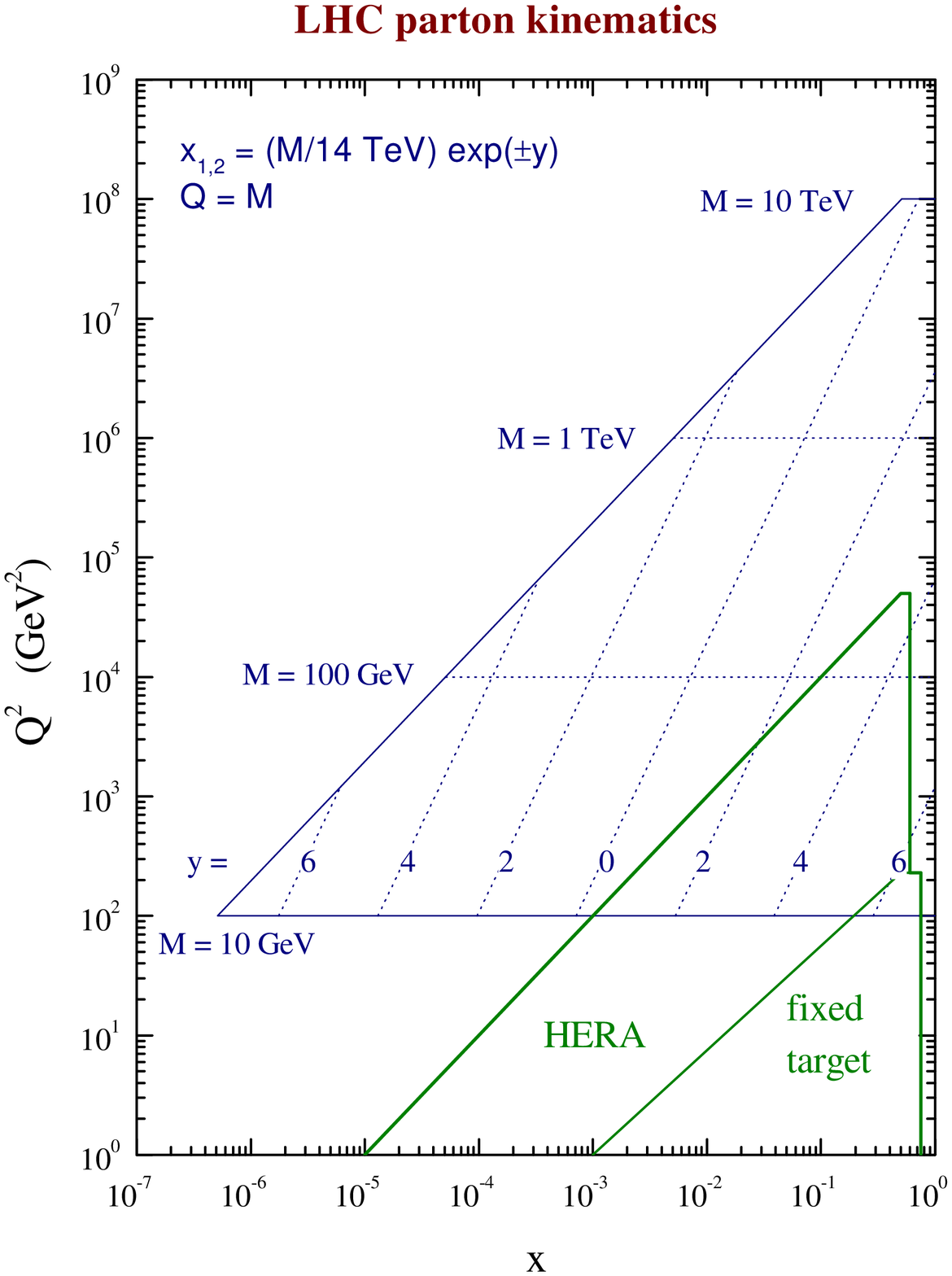,height=2.5in}
\epsfig{file=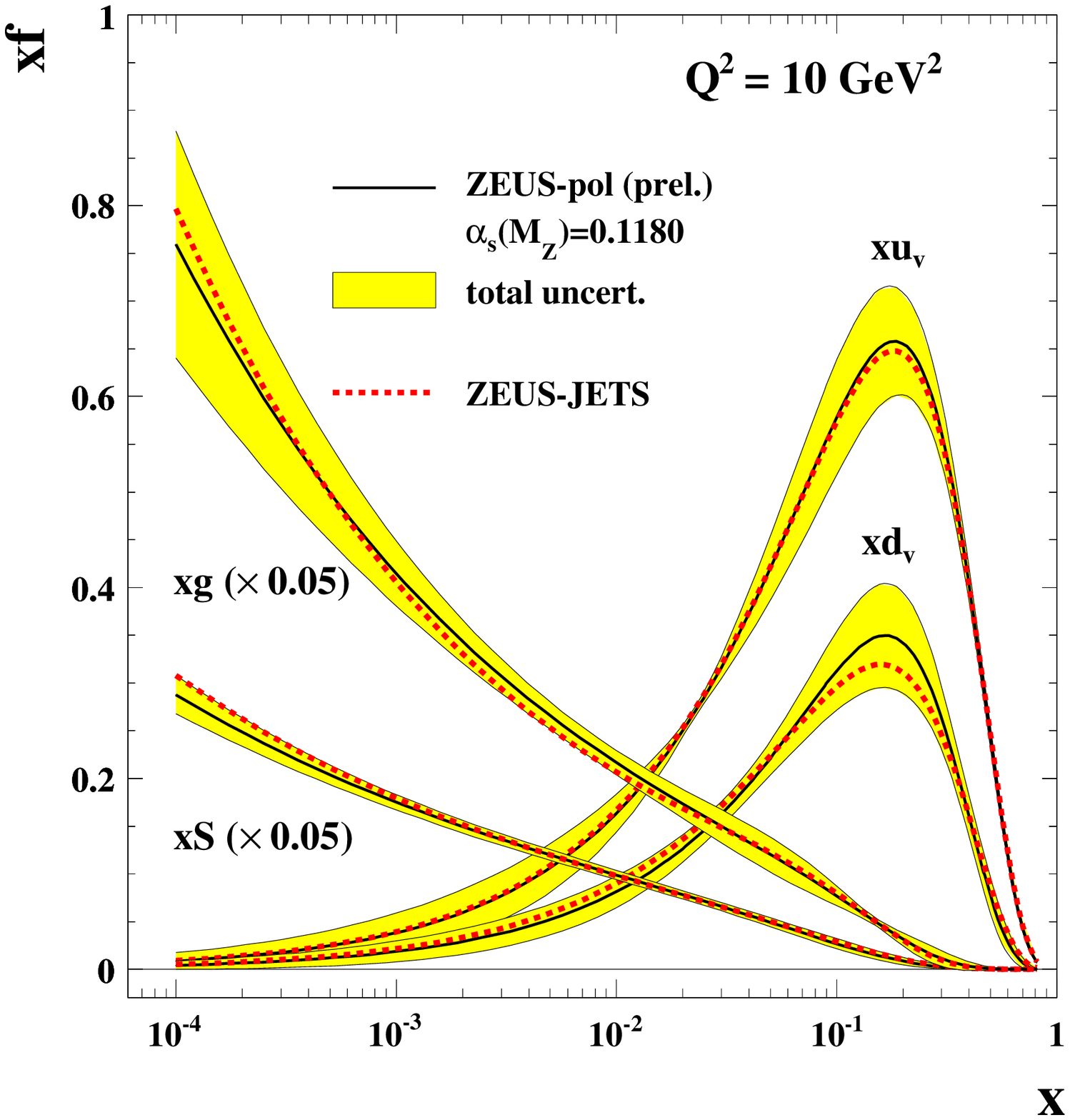,height=2.5in}
\caption{On the left: kinematic plane for the measurement of parton densities at HERA and
other colliders. On the right: example of parton densities extracted from fits to the
ZEUS data. The ZEUS-pol fit, discussed in this report, is shown together with a previous
fit (ZEUS-JETS). The densities for the $u$-valence, $d$-valence, sea and gluon are shown as
a function of $x$ at the scale $Q^2=10~\mathrm{GeV}^2$. The sea and gluon are multiplied
by 1/20 for better visibility.}
\label{fig:kineplane}
\end{center}
\end{figure}

The legacy of H1 and ZEUS is a precise determination of the proton parton densities functions (PDFs),
which is fundamental for the prediction of Standard Model (SM) cross sections at other
colliders, like at the LHC. The region covered by HERA in the kinematic plane in Bjorken $x$
and $Q^2$ is shown in fig.~\ref{fig:kineplane}. Here  $x$ is the fraction of
proton momentum carried by the struck quark and $Q^2$ is
the virtuality of the exchanged boson: the higher the $Q^2$, the deeper we can
look inside the proton. 
At HERA, $x$ spans values between $10^{-5}$ and $10^{-1}$, thus providing
predictions for LHC, which covers similar $x$ ranges but at higher $Q^2$. The values
of $Q^2$ are up to few $\times 10^4~\mathrm{GeV}^{2}$, corresponding to distances inside the
proton smaller than $10^{-18}~\mathrm{m}$. At these values of $Q^2~\simeq M_Z^2,M_W^2$,
HERA can probe DIS processes at the electroweak scale, providing important
results as described in the following. 

The double differential (reduced) cross section for DIS neutral current (NC), 
$e^\pm p \rightarrow e^\pm X$,  can be
written at HERA as:
\begin{equation}
\tilde{\sigma}_r^{\pm}(x,Q^2)  = \frac{d^2 \sigma^\pm}{dx dQ^2}
                     \frac{Q^4x}{2 \pi \alpha^2 Y_+}=
                      \tilde{F}_2^\pm \mp \frac{Y_-}{Y_+}x \tilde{F}_3^\pm - \frac{y^2}{Y_+}\tilde{F}_L,
\end{equation}
where $Y_\pm = 1 \pm (1-y)^2$ and $y$ is the inelasticity of the reaction and is the fraction of the
lepton momentum transferred to the hadronic system in the proton rest frame.
The two proton structure functions $\tilde{F}_2$ and $x\tilde{F}_3$ are discussed in this report in the high-$Q^2$ range. The longitudinal structure function
$\tilde{F}_L$ is only relevant at high $y$ and  is negligible in the kinematic range
reported here. The low-$Q^2$ data constrain the sea and gluon 
at low $x$ (Fig.~\ref{fig:kineplane}), while the neutral current and charged current (CC)
data constrain the $u$-valence and $d$-valence at high-$x$.

\vspace{0.5cm}

\section{Neutral current cross sections at high $Q^2$}


Neutral current events are characterized by the presence of the scattered $e^\pm$
detected at large angle in the calorimeters.
The  NC structure functions can be related to the quark distributions in leading order as:
\begin{eqnarray}
\tilde{F_2}^\pm & = & F_2 + k_Z(-v_e \mp P a_e) \cdot F_2^{\gamma Z} + 
                     k_Z^2(v_e^2+a_e^2 \pm 2 P v_e a_e) \cdot F_2^Z \\
x \tilde{F}_3^\pm &  = & k_Z(\pm a_e + P v_e) \cdot xF_3^{\gamma Z} +
                    k_Z^2(\mp 2 v_e a_e - P(v_e^2+a_e^2)) \cdot xF_3^Z\\
(F_2,F_2^{\gamma Z},F_2^Z) & = & x \sum (e^2_q, 2 e_q v_q, v_q^2+a_q^2)(q + \bar{q}) \\
(xF_3^{\gamma Z},x F_3^Z)&  = & 2x \sum(e_q a_q, v_q a_q) (q - \bar{q}),
\end{eqnarray}
where $k_Z= 1/4 (\sin^2\theta_W \cos^2\theta_W)\cdot (Q^2)/(Q^2+M_Z^2)$,
$a_e = -1/2$ and $v_e=-1/2+2 \sin^2 \theta_W \simeq 0.04$.

In the equations above, $v_i$ and $a_i$ denote the vector and axial weak couplings
of the fermions ($e$ and quarks $q$) to the $Z$, while $P$ is the longitudinal
polarisation of the lepton beam. 
In the expression for $\tilde{F}_2$ (Eq. (2)), the first term 
corresponds to pure $\gamma$ exchange
and dominates at low and medium $Q^2$. At high $Q^2$ the second term, corresponding
to the $\gamma/Z$ interference, becomes relevant, while the third one, due to pure
$Z$ exchange, is always small. The parity violating structure function $x\tilde{F}_3$ 
(Eq. (3))
is dominated by the $\gamma/Z$ interference.

\begin{figure}[htb]
\begin{center}
\epsfig{file=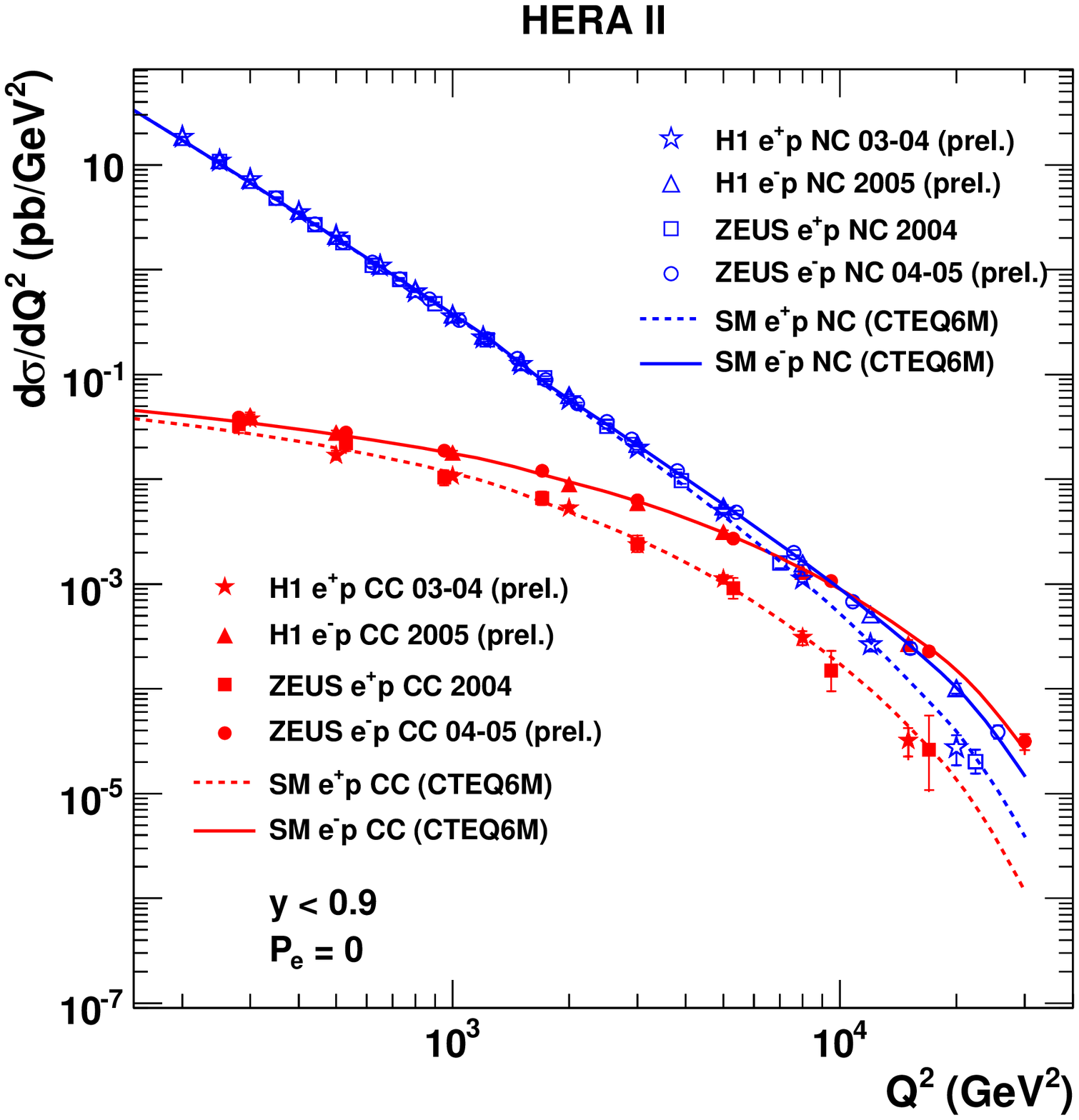,height=2.5in}
\epsfig{file=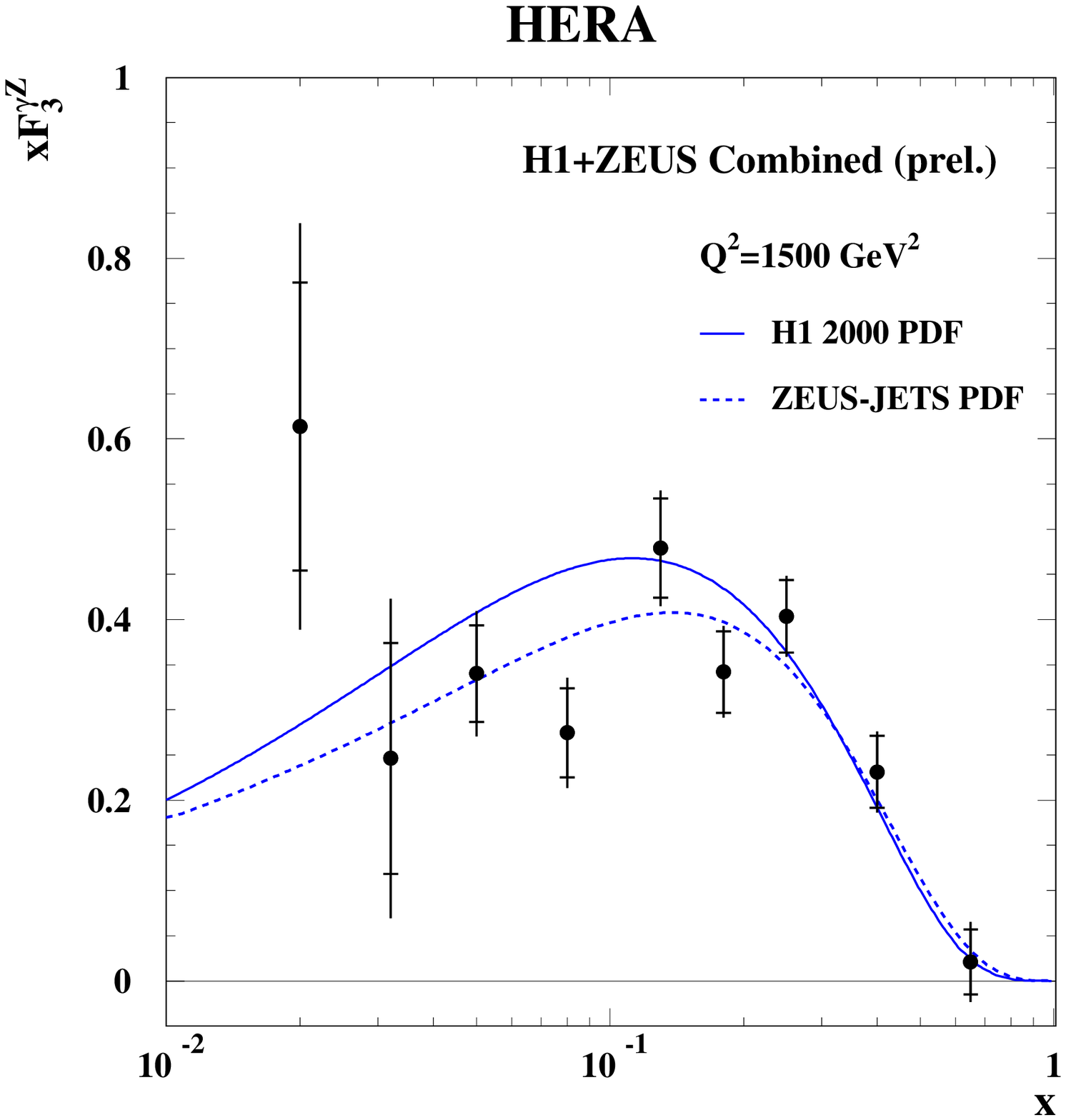,height=2.5in}
\caption{
Results on NC and CC cross sections at HERA. The left plot shows
the differential cross section in $Q^2$ for both reactions. The cross
sections are corrected here to polarisation $P=0$. The right plot
shows the combined H1+ZEUS results on the NC structure function $x F_3$.
The data of the two Collaborations (total of $480~\mathrm{pb}^{-1}$) 
have been here averaged and 
adjusted to a fixed $Q^2$ value of
$1500~\mathrm{GeV}^2$.}
\label{fig:ncresults}
\end{center}
\end{figure}

The unpolarised (corrected to $P=0$) $d\sigma/dQ^2$ cross section 
is shown in Fig.~\ref{fig:ncresults} 
where H1 and ZEUS results
from $e^\pm p$  HERA II data are shown. The cross section is measured over
six orders of magnitude and agrees well with the SM prediction, which is shown 
here using the CTEQ6M PDFs.
The positron and electron cross sections are equal at low $Q^2$ where the photon
exchange dominates, and start to differ at $Q^2 \simeq M_Z^2$ where the $\gamma/Z$ interference
and the contribution of $x F_3$ become sizeable.

As can be seen from Eq. (1),
the parity violating structure function $x \tilde{F}_3$ can be extracted from the difference
between the $e^+ p$ and $e^- p$ cross sections. The difference is dominated by
the interference part which, as $v_e$ is small, is proportional to $a_e a_q$ and 
a parity conserving quantity. In terms of quark densities it can be formulated as
\begin{equation}
xF_3^{\gamma Z}= x/3 (2 u_v + d_v + \Delta),
\end{equation}
(where $\Delta$ is a small correction) and thus gives information on  
$u_v$, $d_v$, the valence distributions for
the up and down quarks. The quantity is shown, for a combination of H1 and
ZEUS data~\cite{ewcombined}, in Fig.~\ref{fig:ncresults}, compared to two different parton
density parametrizations. For the first time, the valence distribution is measured down to
$x~\simeq 0.01$. 
Note that LHC will measure the valence densities down to $x \simeq 0.005$ in
$W$-production.

\section{Charged current cross section}

Charged current events ($ep \rightarrow \nu X$) 
are characterized by the presence of a neutrino
in the final state and are therefore selected requiring high missing transverse
momentum. The differential cross section in $Q^2$ can be written as
\begin{eqnarray}
\frac{d\sigma^{e^+ p}}{dx,dQ^2}= (1+P) \frac{G_F}{2 \pi}
                      \frac{(M^2_W)^2}{(M^2_W+Q^2)^2 } 
                      [ \bar{u}_i(x,Q^2)+(1-y)^2 d_i(x,Q^2)] 
\label{eqcc1} \\
\frac{d\sigma^{e^- p}}{dx,dQ^2}= (1-P) \frac{G_F}{2 \pi}
                     \frac{( M^2_W)^2}{(M^2_W+Q^2)^2} 
                      [ u_i(x,Q^2)+(1-y)^2 \bar{d}_i(x,Q^2)].
\label{eqcc2}
\end{eqnarray}

The cross sections for $e^+$ and $e^-$ have different dependences
on the $u$-type and $d$-type parton densities, providing an important
flavour separation at high $x$. As shown in Fig.~\ref{fig:ncresults},
the CC cross section for $e^- p$ is higher
than that of $e^+p$ due to the fact that the $u$ density is larger
and in addition the $d$ density in Eq.~(\ref{eqcc1}) is suppressed by the helicity
factor $(1-y)^2$. The d$\sigma/dQ^2$ is flat for $Q^2 \ll M_W^2$, while for
$Q^2 > \sim M_W^2$  it decreases and its size becomes similar to that of the NC
interaction. This text-book plot is one of the confirmation of the
unification of the electromagnetic and weak interactions at the
scale of the masses of the $W$ or $Z$. 

The term $(1 \pm P)$ in Eqs. (\ref{eqcc1}) and (\ref{eqcc2}) refers to the
strong dependence of the CC cross section on the lepton-beam polarisation. 
At HERA, leptons became naturally transversely polarised through synchroton
radiation
via the Sokolov-Ternov effect.
 Spin rotators
were installed during the luminosity upgrade in order to provide 
longitudinally polarised beams for the experiments.  
The build-up time for the polarisation was approximately  30 minutes at the
beginning of the machine fills.
The polarisation,
here defined as $P=(N_R-N_L)/(N_R+N_L)$,
where $N_R$ ($N_L$) is the number of right-handed (left-handed) electrons,
reached typically values of $30$-$40\%$ and its sign was changed every 
two-three months. It was measured by three different devices, exploiting the
dependence on $P$ of the Compton scattering cross section of 
circularly-polarised photons with the lepton beam.

\begin{figure}[htb]
\begin{center}
\epsfig{file=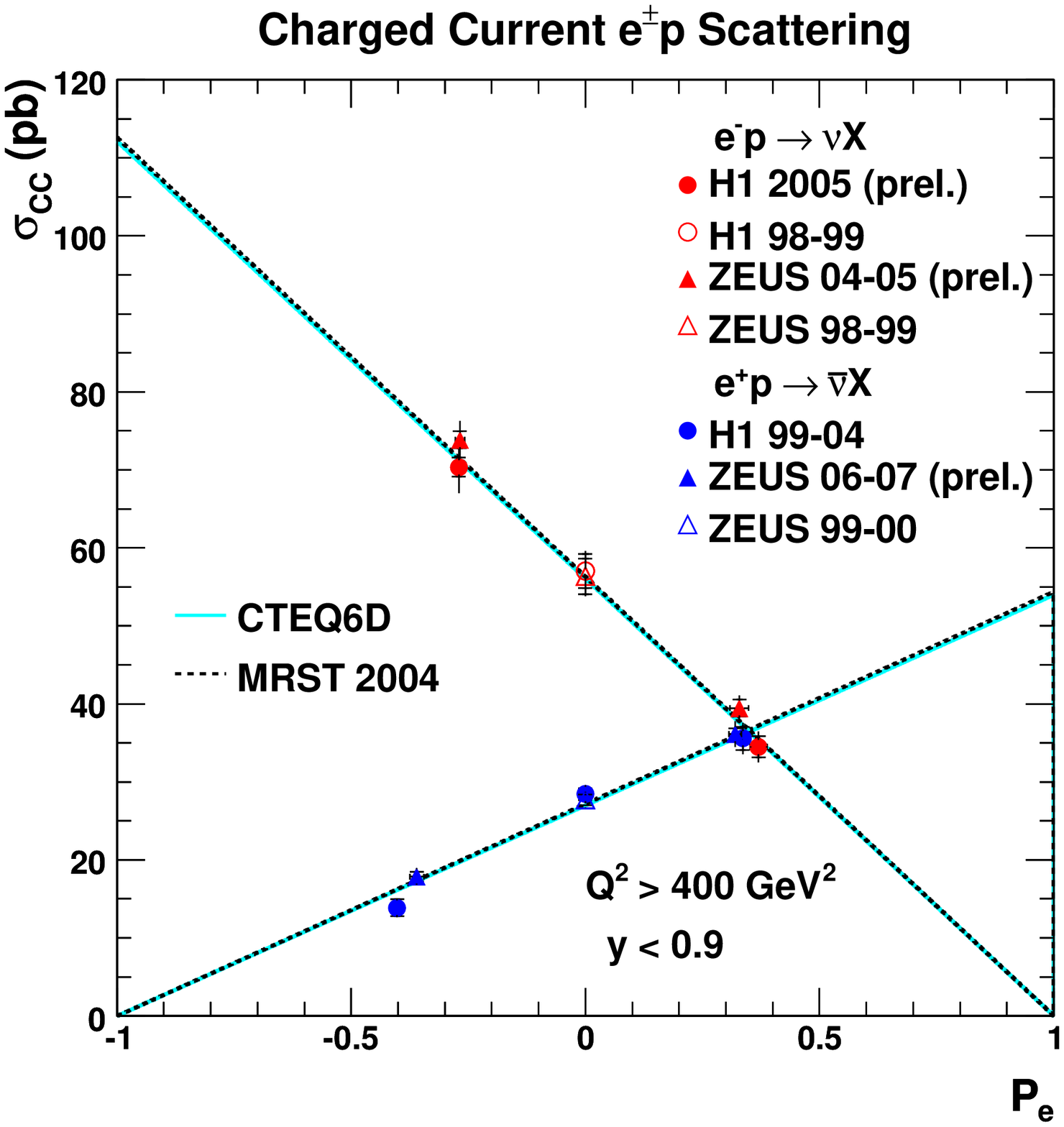,height=2.5in}
\epsfig{file=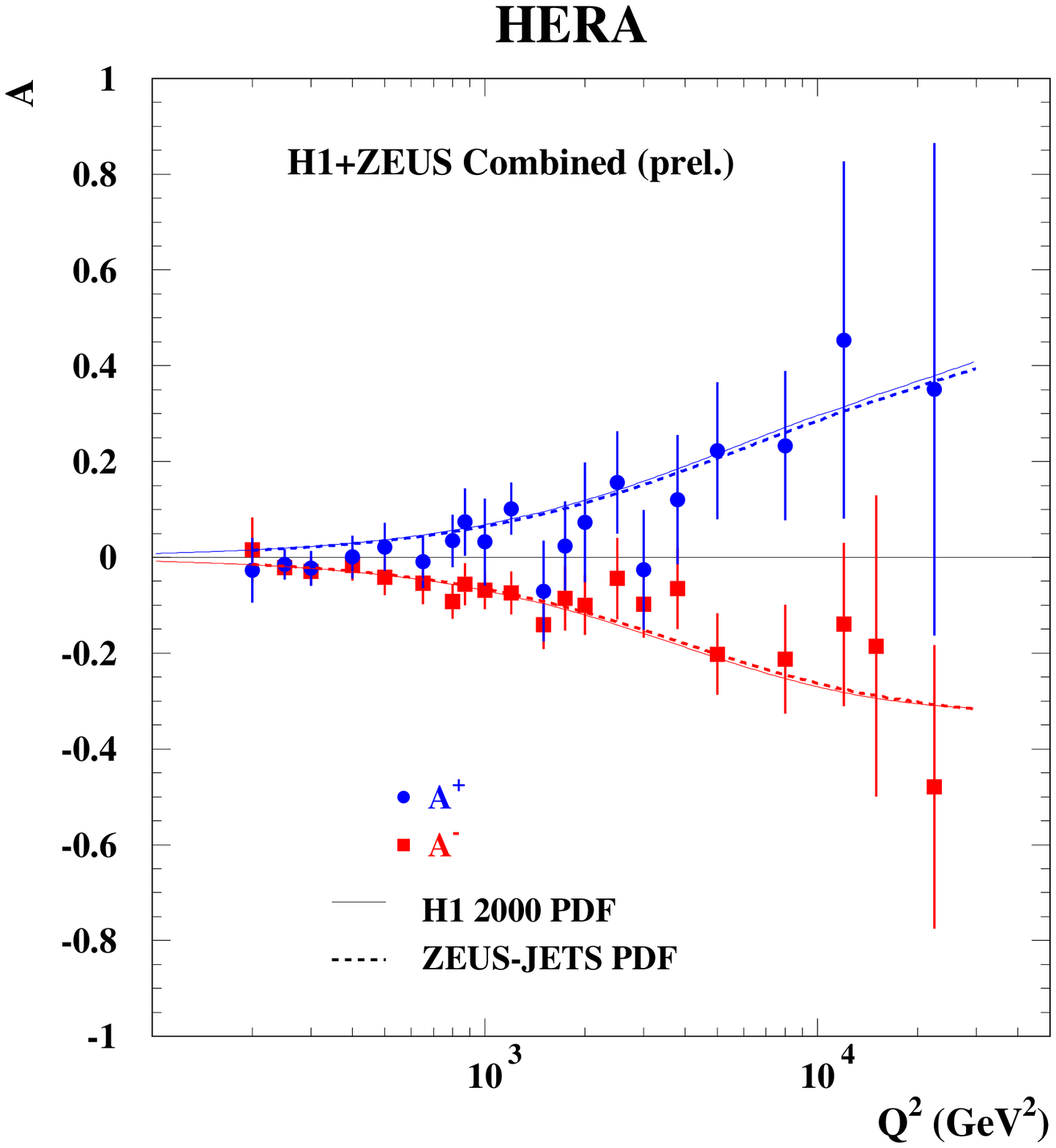,height=2.5in}
\caption{
Results on CC and NC polarised cross sections at HERA. The left plot shows
the total cross section for charged current $e^\pm p$ versus the lepton
beam polarisation. 
The right plot
shows the combined H1+ZEUS results for the asymmetry in NC interactions.}
\label{fig:pol}
\end{center}
\end{figure}

The linear dependence of the total CC cross section versus $P$ is shown in
 Fig.~\ref{fig:pol}, another text-book plot, which shows that the
$e^-p$ ($e^+ p$) cross sections tends to zero for $P=+1$ ($P=-1$). This
is expected from the absence of right-handed currents in the weak interaction.

\section{Electroweak couplings}

The effect of the polarisation in NC is much weaker than in CC and visible only
at very high $Q^2$. In order to enhance it, the asymmetry
\begin{equation}
A^\pm= \frac{2}{P_R-P_L} \cdot
       \frac{\sigma^\pm(P_R)-\sigma^\pm(P_L)}
            {\sigma^\pm(P_R)+\sigma^\pm(P_L)}\simeq 
            \mp k_Z   a_e \frac{F_2^{\gamma Z}}{F_2}
\end{equation}
is measured. This quantity is proportional to combinations $a_e v_q$, thus it provides
a direct measurement of parity violation in neutral current at high $Q^2$. The
asymmetry is shown in Fig.~\ref{fig:pol}, where it can be seen 
that $A^+$ is positive
and $\simeq -A^-$, as predicted by the theory~\cite{ewcombined}. This asymmetry
is the first measurement of parity violating effects in weak interactions at
distances down to $10^{-18}~\mathrm{m}$. 

\begin{figure}[htb]
\begin{center}
\epsfig{file=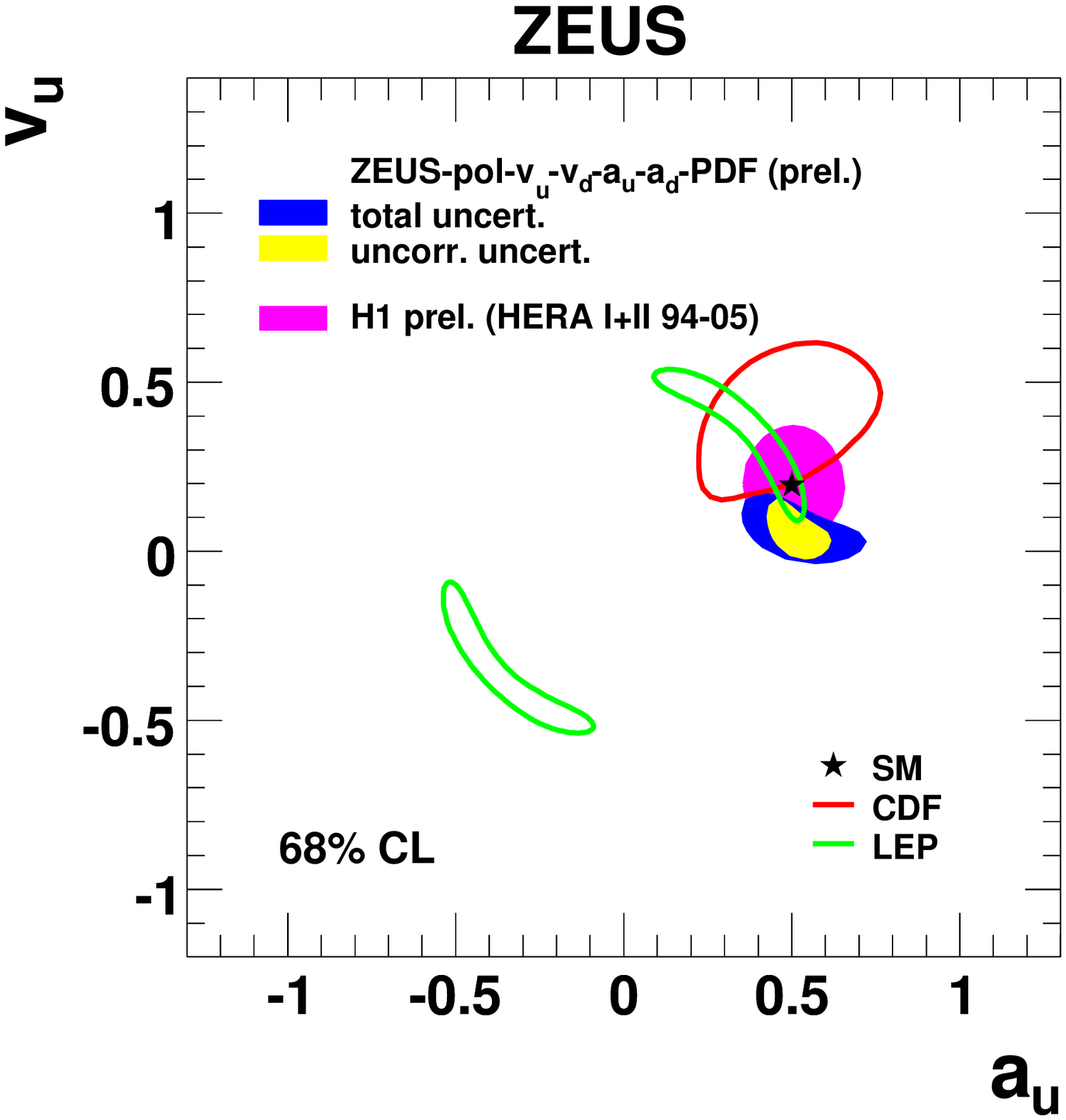,height=2.7in}
\epsfig{file=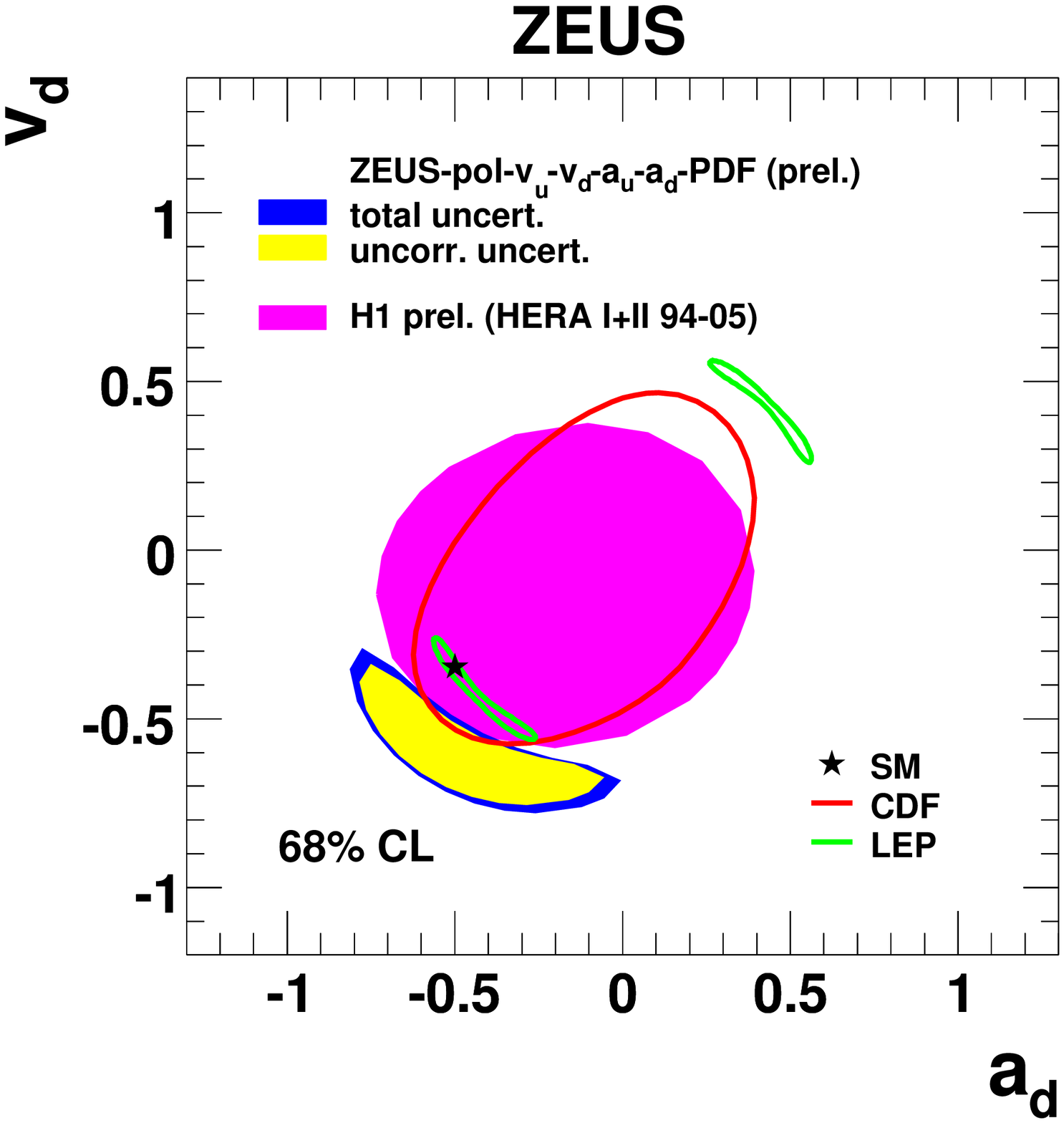,height=2.7in}
\caption{Contour plots at the $68\%$ confidence level ($CL$) of the limits on the electroweak
couplings of the $u$ (left) and $d$ (right) quarks to the $Z$. The H1 and ZEUS results
are shown together with CDF and LEP contours.}
\label{fig:ewcouplings}
\end{center}
\end{figure}

The neutral current polarised data can also be employed to perform
a QCD fit and extract simultaneously the parton densities and the electroweak 
parameters. The next-to-leading-order (NLO) QCD fits 
parametrise the parton densities at an initial scale $Q_0^2$ and evolve them in $Q^2$
with the DGLAP evolution equations. The new polarised NC data from HERA II provide
a better constraint on the $u$-density, thanks to the much
higher statistics ($\simeq 176~\mathrm{pb}^{-1}$) of 
the $e^-p$ sample compared to HERA I ($\simeq 16~\mathrm{pb}^{-1}$).
In addition, the polarisation allows the determination of the axial and vector
couplings of the $u$ and $d$ quarks to the $Z$, which are still not
very well constrained by other colliders data.

From the Eqs. (2-5), as $v_e$ and $k_Z^2$ are very small, it can be seen that the
unpolarised term in $x F_3$ is sensitive to the product $e_q a_e a_q$ and
thus to the axial coupling of the quarks. The polarised term in
$F_2$, instead, is sensitive to the product $e_q v_q a_e$ and thus to the
vector coupling to the quarks. The new ZEUS-pol fit~\cite{zeuspol} provides information
on the parton densities (Fig.~\ref{fig:kineplane}) and on the quark couplings at
the same time. The quark couplings are shown in Fig.~\ref{fig:ewcouplings}, together
with results from CDF and LEP II. The H1 results are also shown.
The precision of the HERA data is competitive and the sign ambiguity of
the LEP II data is resolved. All four couplings are in good agreement with
the SM expectation.

\begin{figure}[htb]
\begin{center}
\epsfig{file=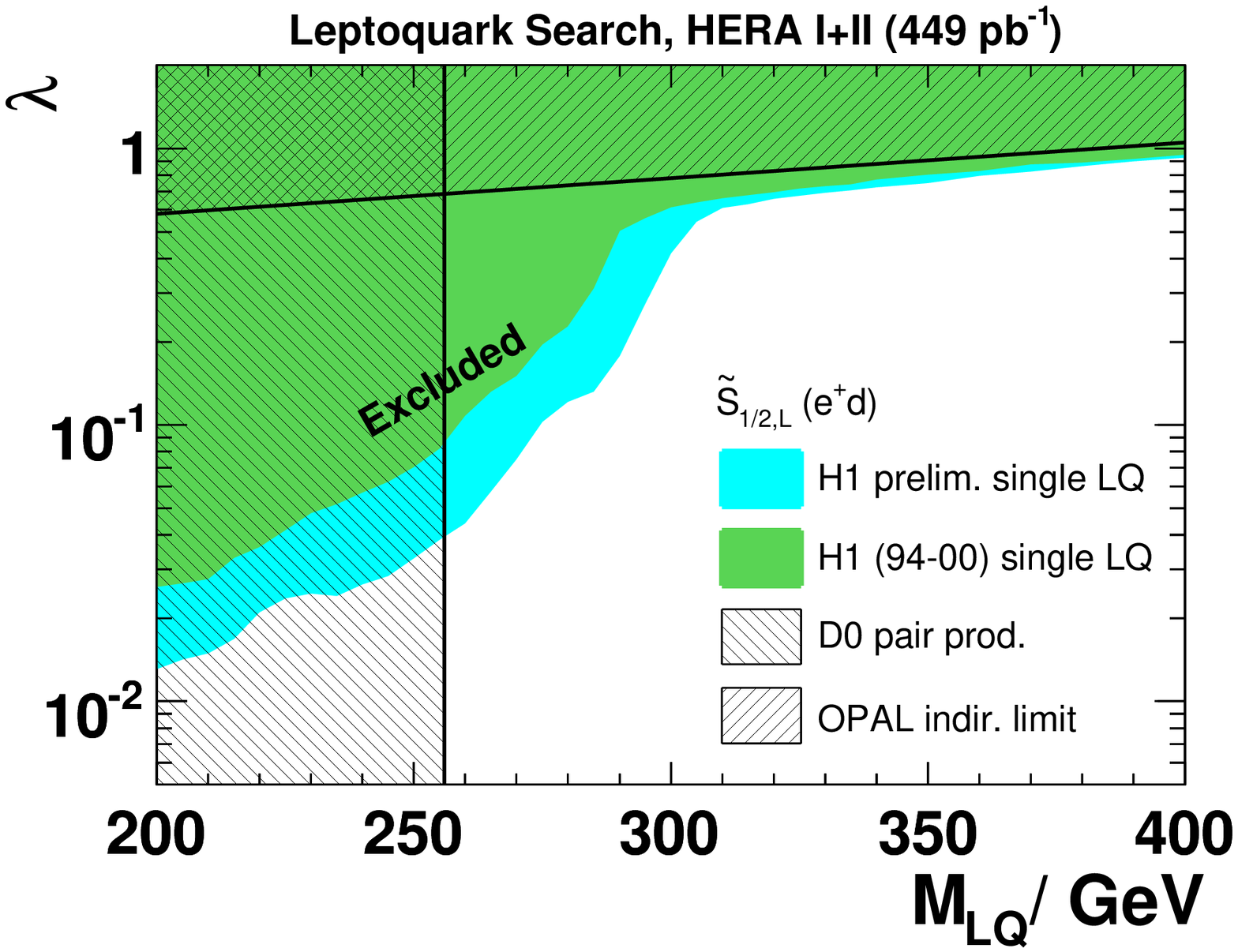,width=2.7in}
\epsfig{file=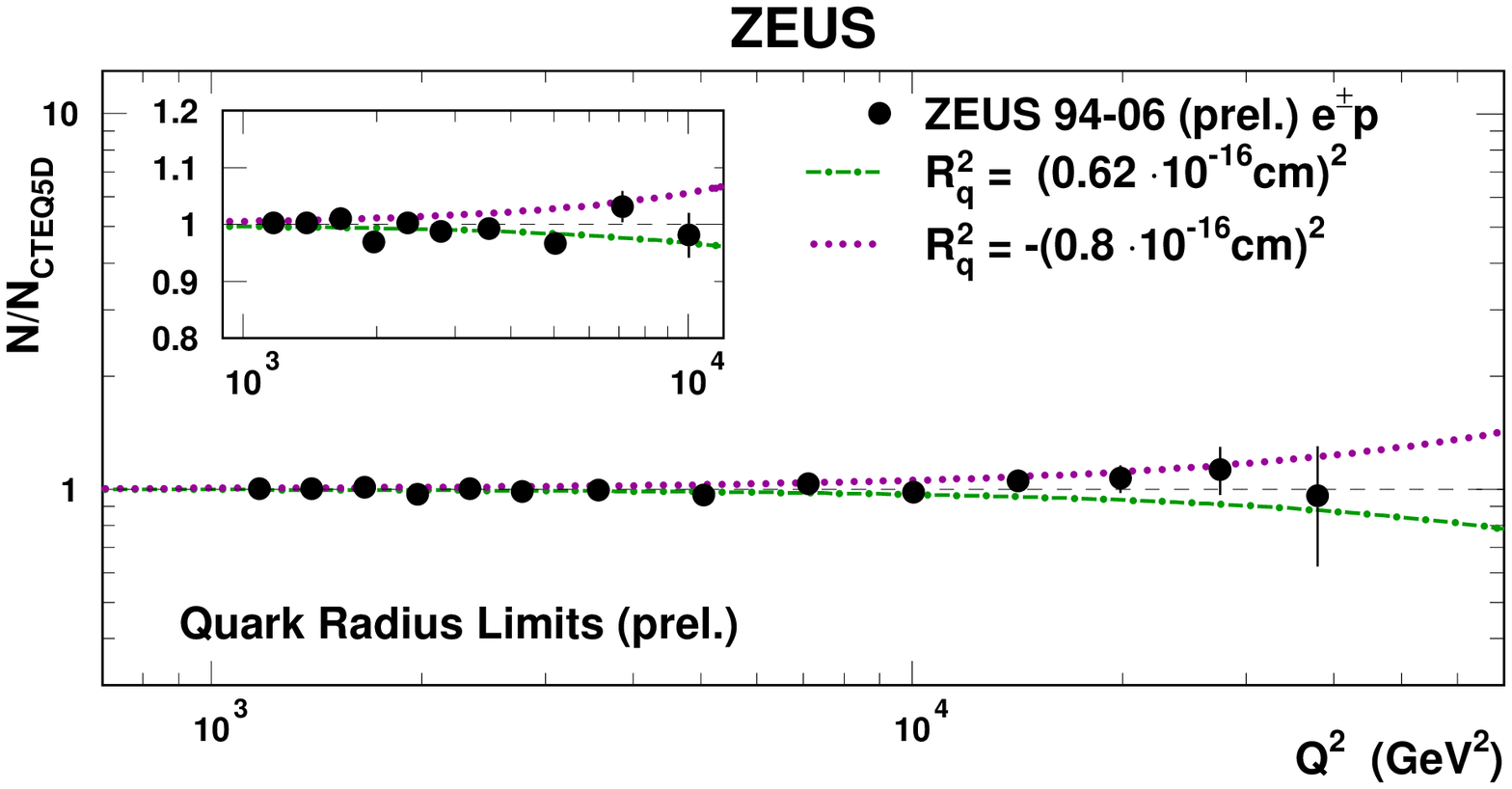,width=2.7in}
\caption{On the left: example of limit on one 
of the leptoquark types from the H1 data,
based on the HERA I+II statistics.  The H1 exclusion limit is shown for
the leptoquark $\tilde{S}_{1/2,L}(e^+ d)$, as a function of its mass and its
Yukawa coupling $\lambda$ to the positron and quark. 
On the right:
Limit on the quark radius from the 1994-2006 ZEUS data. 
The differential cross section versus $Q^2$ is shown divided by the 
SM expectation using CTEQ5D, a parametrisation which does
not include the latest HERA data. The line shows the expectation for a quark radius
corresponding to the obtained limit at 95$\%~CL$.}
\label{fig:contact}
\end{center}
\end{figure}

\section{Search for new physics at high $Q^2$}

The HERA data at high $Q^2$ have triggered the search for new physics in
$eq$ interactions down to distances of $10^{-18}~\mathrm{m}$.  
A classical search is that for leptoquarks
(LQs) coupling to
first generation fermions, that can be produced
in the $s-$ and $u-$channels from the fusion of the lepton beam with
one of the valence quarks in the protons. The leptoquarks decay 
then to an electron/positron and 
a quark, or to a neutrino and a quark, with a signature similar to
that of DIS NC or CC events at high $Q^2$. 
For masses of the leptoquark $M_{LQ} < \sqrt{s}$,
the LQ signal appears as a narrow resonance at a certain Bjorken
$x \simeq M^2_{LQ}/s$. For masses greater than $\sqrt{s}$, it manifests
itself as a contact interaction, causing deviations of the cross sections 
at high $x$ and $Q^2$ from the SM expectation.
The LQ signal can be
distinguished also thanks to
the different angular distribution of the lepton and jet in the final state
in the LQ rest frame, compared to that of the DIS background.  
H1 has recently presented a search in the whole statistics 
from HERA I+II ($\simeq 450~\mathrm{pb}^{-1}$)~\cite{h1lq}, 
but no deviation from the SM was observed in the final state lepton-jet mass spectra.
Limits were set on LQ production, an example is shown in Fig.~\ref{fig:contact}, together with recent results
from other colliders. For masses in the range between 260 and 300 GeV, 
HERA excludes the production of leptoquarks with LQ-fermions Yukawa  couplings in the range
$10^{-2}$-$10^{-1}$, in a region not covered by other experiments.

The substructure of quarks is another classical type 
of new physics that could be
visible in the HERA data. It would manifest itself as a contact interaction,
which decreases the $Q^2$ spectrum with a form factor of the form
$(1-R^2_q/6 \cdot Q^2)^2$, where $R_q$ is the finite quark radius and
the electron is assumed to be pointlike. The $d\sigma/dQ^2$ spectrum, normalized
to the theory, is shown for the ZEUS 1994-2006 data~\cite{zeusci}
 in Fig.~\ref{fig:contact}.
The ratio is consistent with one, thus no evident sign of a quark
substructure is seen in these data. A limit of $R_q< 0.62 \times 10^{-18}~\mathrm{m}$
is obtained at $95\%~CL$. This is the most stringent
limit to date, and not very far from the electron radius limit~\cite{bourilkov}
determined at LEP.

\section{Constraints from events with high $p_T$ leptons}

The observation of particular events in H1, with high transverse-momentum ($P_T^l$) leptons
in the final state, has attracted quite some attention from the HERA community
for many years and triggered investigations of possible new physics. 

The first search was the observation by H1 in 1994 of an event with an isolated high-$P_T^l$ muon,
large missing transverse momentum ($P_T^\mathrm{miss}$) and a jet with large transverse
energy $P_T^X$. Since then H1 has reported regularly on the observation of
an excess of this type of events, both with an electron/positron or a muon in the
final state, exclusively in  $e^+p$ interactions. In the Standard Model, such events are due
to $W$ production, $ep \rightarrow (e)WX \rightarrow (e) l \nu X$, where the $W$ is radiated
from one of the quarks in the proton. The $W$ can decay then to a lepton and neutrino, while
the struck quark produces the hadronic final state $X$. The scattered electron/positron can be 
seen in the detector if it is scattered at large angle (for $Q^2$ typically $>4~\mathrm{GeV^2}$) 
or escapes in the beam pipe in the lepton-beam direction.
In $W$ production, however, the hadronic system $X$ has typically a low transverse
energy, while the events observed by H1 showed an excess over the prediction at
$P_T^X>25~\mathrm{GeV}$.  
The ZEUS Collaboration has never confirmed this excess and recently the two
Collaborations have defined a common phase space to compare the results~\cite{isolleptons}.
The final state leptons are restricted to a common acceptance region corresponding
to the polar-angle range $15^\circ < \theta^l < 120^\circ$ (where $\theta^l$ is measured
with respect to the proton-beam direction) and $P_T^l>10~\mathrm{GeV}$. The cut on
the missing transverse energy of the event was
chosen at $P_T^\mathrm{miss}>12~\mathrm{GeV}$. The full HERA I+II data set was combined
($0.97~\mathrm{fb}^{-1}$) and in total 64 electron events and 23 muon events were
selected, in good agreement with the SM expectation of $72.9\pm 8.9$ and $19.9 \pm 2.6$,
respectively.

\begin{table}[b]
\begin{center}
\begin{tabular}{l|cc}  
$e^+ p, ~P_T^X>25~\mathrm{GeV}$      &  $e+\mu$ Data &  $e+\mu$ SM  
\\ \hline
H1  $0.29~\mathrm{fb}^{-1}$      &  17           &  $7.1 \pm 0.9$  \\
ZEUS $0.29~\mathrm{fb}^{-1}$     &   6           &  $7.5 \pm 1.1$  \\ \hline
H1+ZEUS $0.58~\mathrm{fb}^{-1}$  &   23          &  $14.6 \pm 1.9$  \\ \hline
\end{tabular}
\caption{Data yields for the isolated lepton events at high $P_T^X$ in the full $e^+p$
H1+ZEUS data set. The selection refers to the common acceptance region in the two
detectors.}
\label{tab:isolleptons}
\end{center}
\end{table}

\begin{figure}[htb]
\begin{center}
\epsfig{file=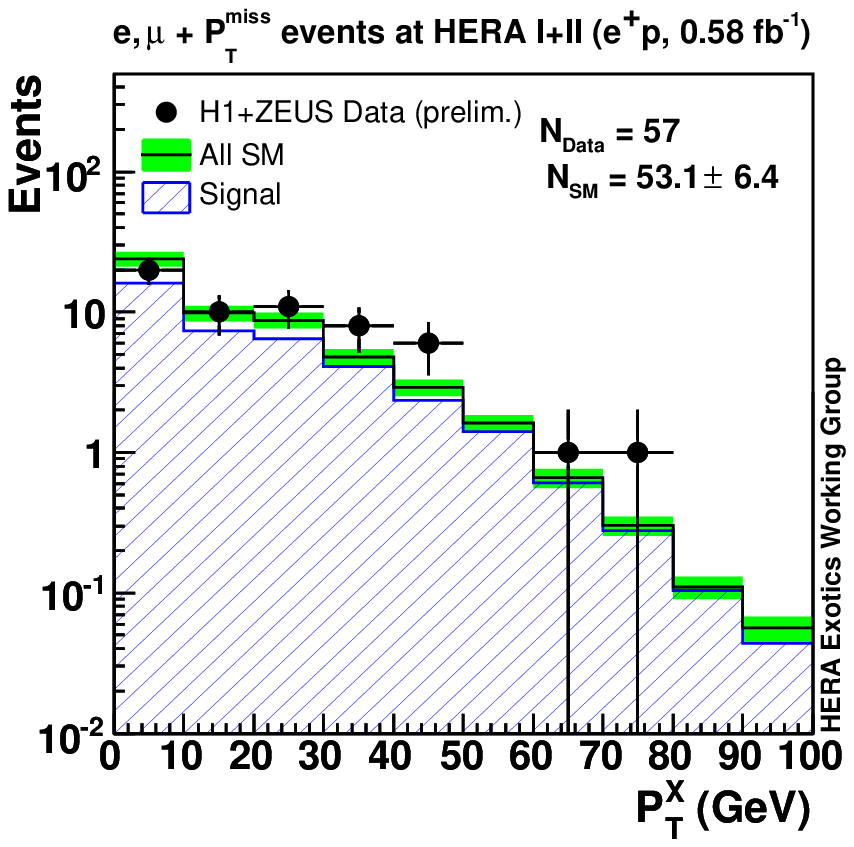,height=2.5in}
\epsfig{file=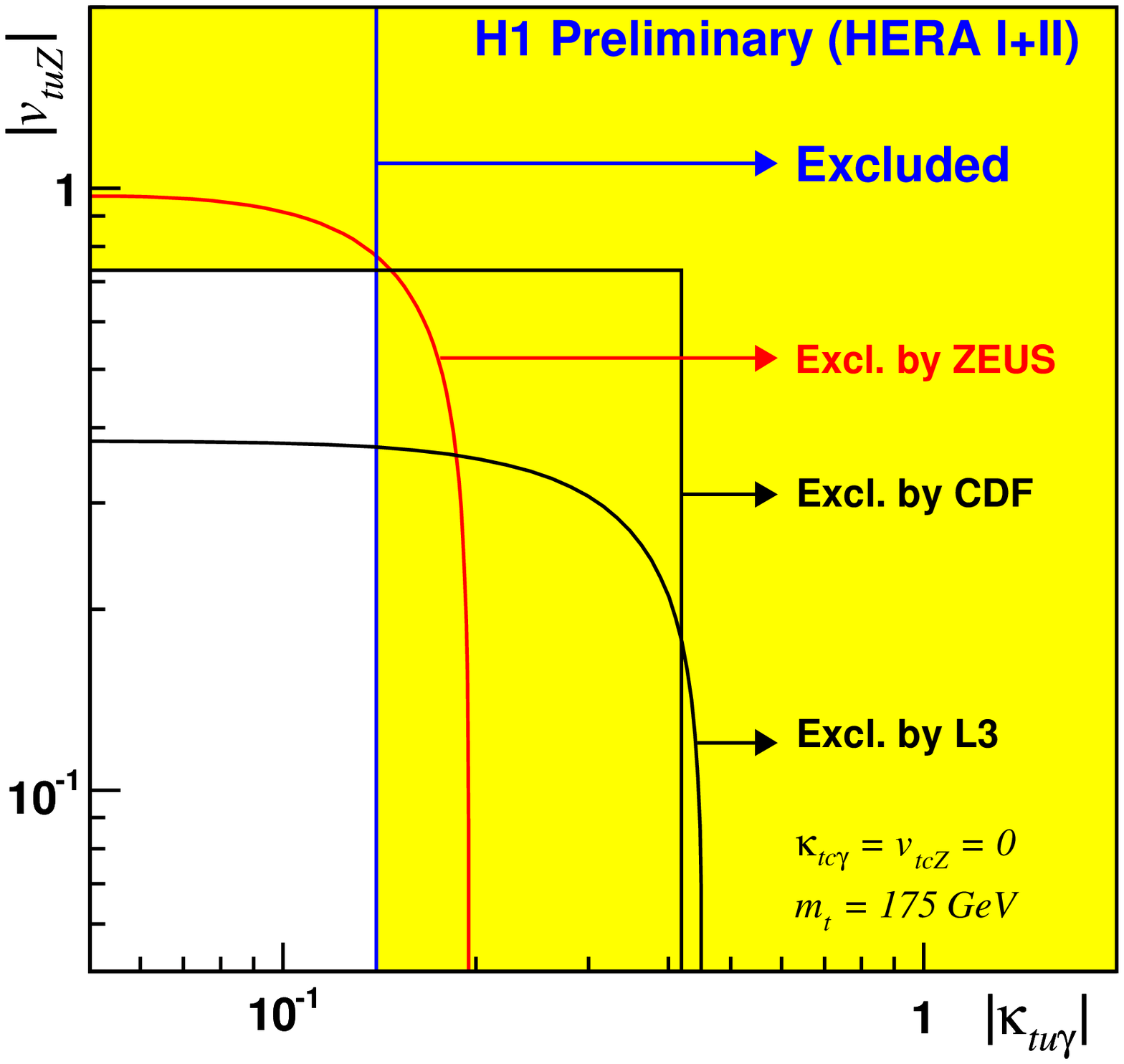,height=2.5in}
\caption{On the left: distribution of the $P_T^X$ in the HERA combined $e^+ p$ data. 
On the right: Limits on the FCNC coupling for single top production from the
latest H1 data, compared to previous published results. The contribution from charm
quark is negligible in this kinematic regime at HERA and it is therefore neglected
in the extraction of the limit ($k_{t c \gamma}=0$). The couplings to the $Z$ are also
neglected in the latest H1 results.}
\label{fig:isol}
\end{center}
\end{figure}

\begin{figure}[htb]
\begin{center}
\epsfig{file=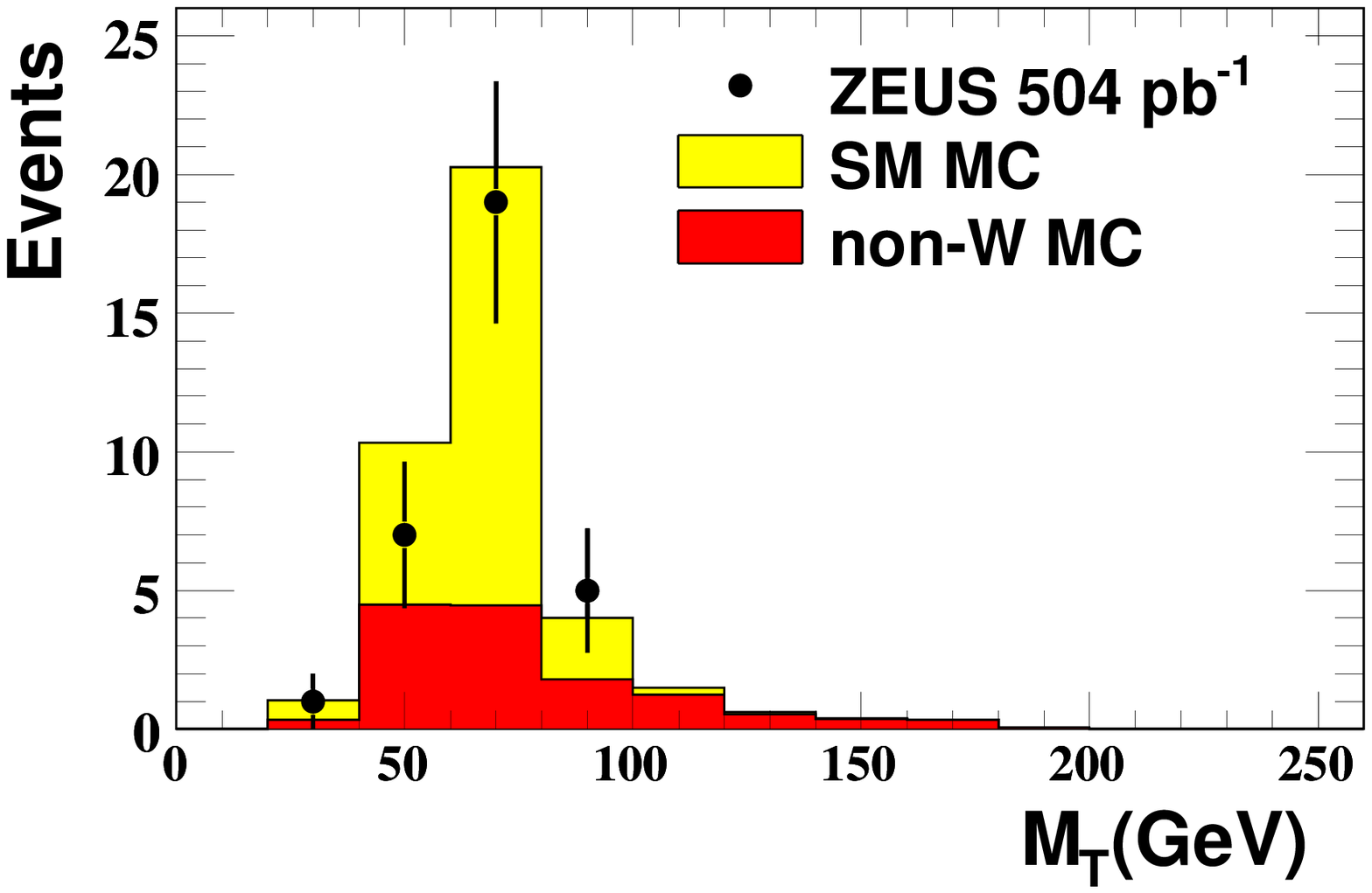,width=2.7in}
\epsfig{file=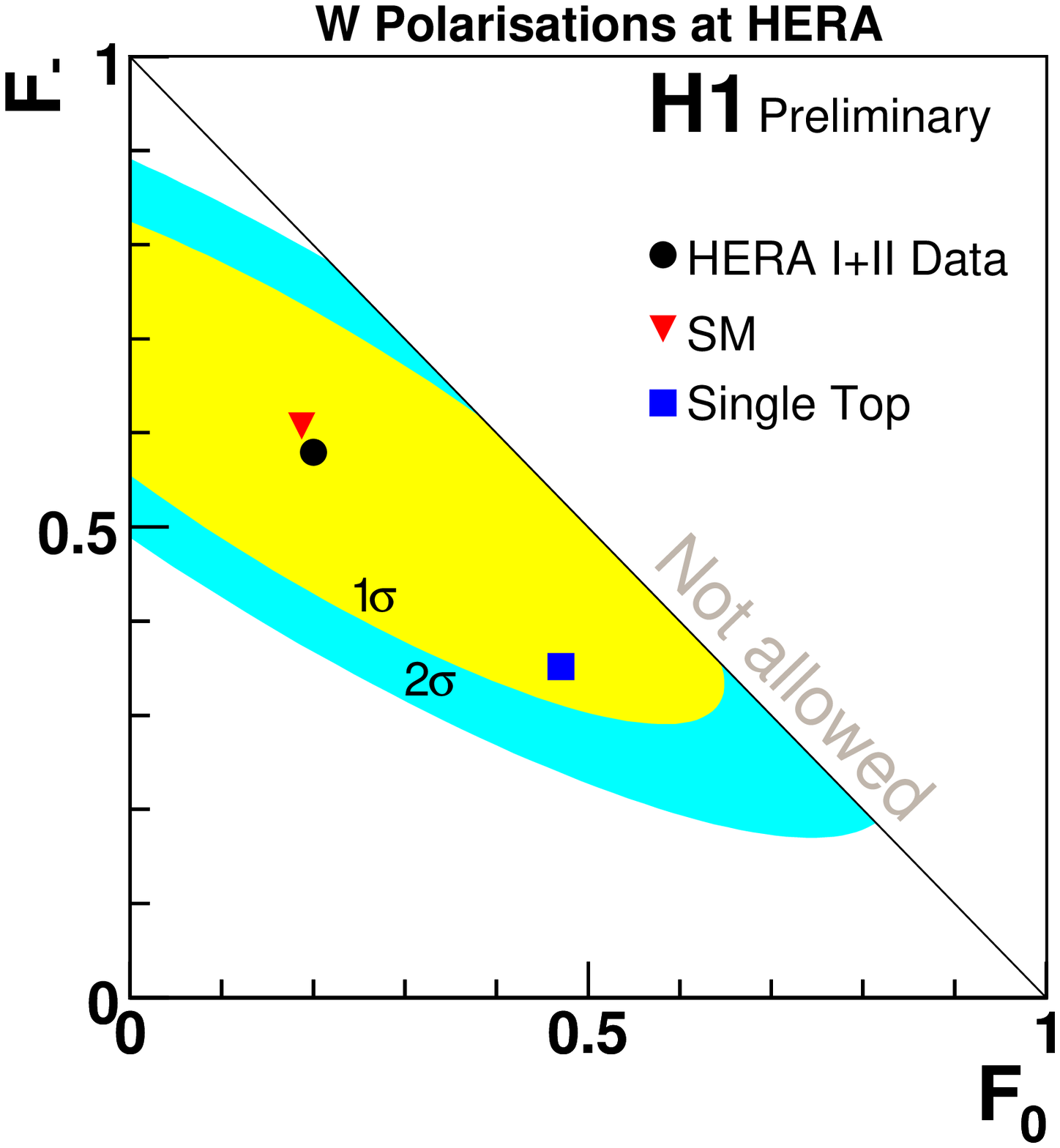,height=2.0in}
\caption{On the left:
distribution of the transverse mass in $W$-candidate events in ZEUS.
The data (dots) are superimposed to the SM expectation which is dominated
by $W$ production (light shaded histogram). On the right: Contour plots at
$1 \sigma$ and $2\sigma$ for the fitted $F_-$ and $F_0$ polarisation
fractions of the $W$, from the H1 data.}
\label{fig:wcross}
\end{center}
\end{figure}

The combined $P_T^X$ spectrum is shown in Fig.~\ref{fig:isol} for the $e^+p$ data only. 
Selecting the region at high $P_T^X$, H1 observed 17 events with 
$P_T^X>25~\mathrm{GeV}$, compared to a SM expectation of $7.1 \pm 0.9$, corresponding to
a $2.9\sigma$ deviation. The ZEUS data yield is shown in Table~\ref{tab:isolleptons}. No
excess is observed by ZEUS and when the two data sets are combined, a total of
$23$ events is observed, for a SM prediction of $14.6~\pm 1.9$. The deviation is
reduced to $1.8\sigma$ and it is therefore not significant.

One of the possible processes giving rise to events at high $P_T^X$ is the
production of single top at HERA, $ep \rightarrow (e)tX$, where the top decays
as $t \rightarrow bW \rightarrow b l \nu$. The SM cross section for such process is very small,
of the order of $1~\mathrm{fb}$, therefore the observed event yield could   
be only due to new physics. A possible process is  the 
flavour changing neutral current (FCNC) transition
of a $u$ quark in the proton to a $t$ quark. This anomalous single-top production
is described by an effective Lagrangian where the interaction of
a top with a $u$ quark and the photon is characterized by the magnetic
coupling $k_{t u \gamma}$. H1 analysed their interesting events optimizing
the cuts for single-top search~\cite{h1singletop}. 
The selected events distributions resulted to be compatible
with the SM expectations and a limit at $95\%~CL$ on the cross section
of $\sigma(ep \rightarrow etX) < 0.16~\mathrm{pb}$ was extracted. The corresponding
limit on $k_{t u \gamma}<0.14$ is shown in Fig.~\ref{fig:isol}, where also previous
published results from other colliders are shown. The H1 preliminary limit is
the strictest limit to date.

On the other hand, the events selected by H1 and ZEUS are a good starting
sample to select $W$ events at HERA. The transverse mass in the $W \rightarrow e \nu$ channel
is shown for the ZEUS data in Fig.~\ref{fig:wcross}, where the data clearly show the
reconstructed $W$-mass peak. ZEUS measured a cross section~\cite{zeusw} for
$\sigma(W \rightarrow l \nu X)$ of $0.89^{+0.25}_{-0.22}(\mathrm{stat.}) \pm
                                         0.10(\mathrm{syst.})~\mathrm{pb}$, in good agreement
with the SM calculation of $1.2~\mathrm{pb}$. 

Additional selection cuts were
also applied in H1, selecting a total of 31 $W$ candidates with very high purity. These
events were analysed in terms of the polarisation fractions of the
$W$ boson~\cite{h1w}, which were defined as a function of the angle $\theta^*$ between
the W momentum in the laboratory frame and the charged decay lepton in the rest
frame of the $W$. The left-handed $F_-$, the longitudinal $F_0$ and the right-handed
$F_+$ fractions have to satisfy the relation $F_+ + F_- + F_0=1$ and can be
extracted fitting the measured $\cos\theta^*$ distributions from the relation 
($i.e.$ for $W^+$):
\begin{equation}
\frac{dN}{d \cos\theta^*}  \propto  (F_+) \cdot \frac{3}{8}(1+\cos\theta^*)^2 
                           +        F_0 \cdot \frac{3}{4}(1-\cos^2\theta^*) 
                            + F_- \cdot \frac{3}{8}(1- \cos\theta^*)^2.
\end{equation}    
The optimal values for $F_0$ and $F_-$ extracted from a simultaneous fit are shown
in Fig.~\ref{fig:wcross}, showing values very close to the ones predicted by the
SM, but also compatible with single-top production within $1 \sigma$.

\begin{table}[b]
\begin{center}
\begin{tabular}{l|cc}  
Data sample      &  Data &  SM  \\ \hline
$e^+ p$   ($0.56~\mathrm{fb}^{-1}$)      &  5           &  $1.82 \pm 0.21$  \\
$e^- p$   ($0.38~\mathrm{fb}^{-1}$)      &  1           &  $1.19 \pm 0.14$  \\ \hline
$e^\pm p$ ($0.94~\mathrm{fb}^{-1}$)      &  6           &  $3.00 \pm 0.34$  \\ \hline
\end{tabular}
\caption{Data yields for the $2e$ and $3e$ events 
for $\sum P_T>100~\mathrm{GeV}$ in the full HERA data set.}
\label{tab:multileptons}
\end{center}
\end{table}

\begin{figure}[htb]
\begin{center}
\epsfig{file=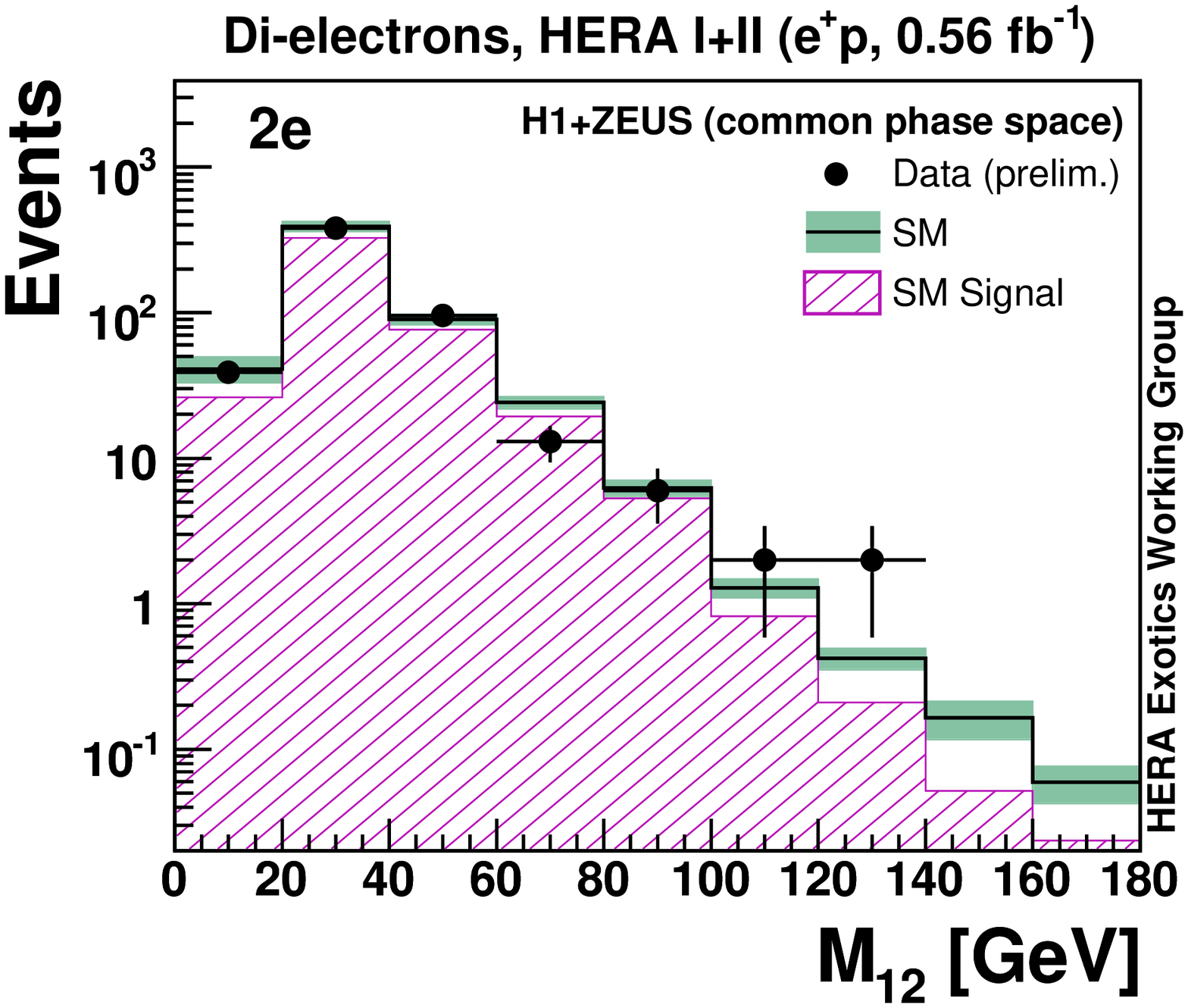,width=2.7in}
\epsfig{file=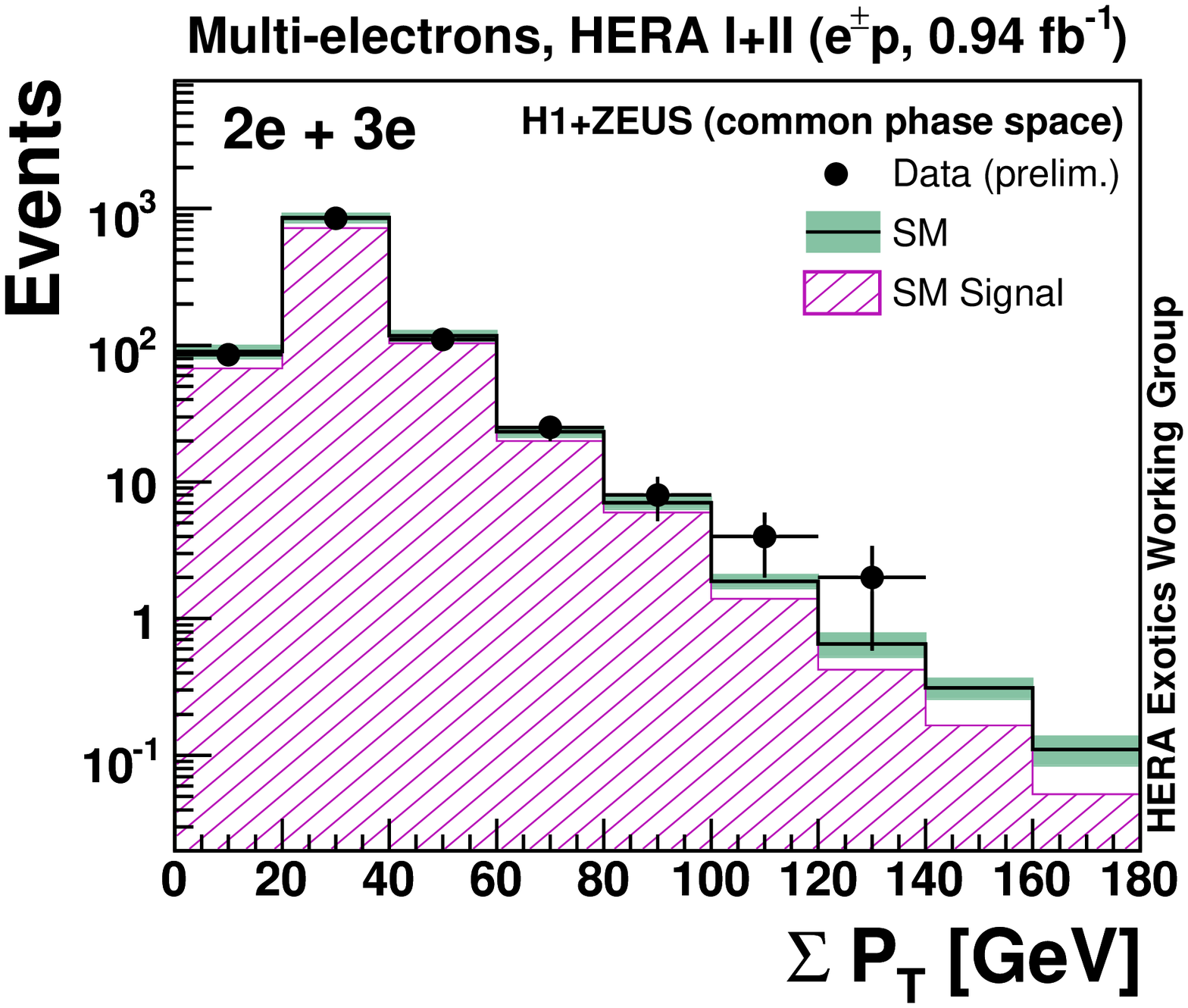,width=2.7in}
\caption{On the left:
distribution of the mass of the two highest-energy leptons in $2e$ events
in the $e^+p$ combined HERA data.
On the right: distribution of the sum of the transverse momentum
in $(2e)$+$(3e)$ events in the full HERA sample.}
\label{fig:multileptons}
\end{center}
\end{figure}

The H1 Collaboration also reported the observation, in the HERA I data, of
outstanding multielectron/positron events, with either two (2$e$) or three ($3e$) 
isolated high-$P_T^l$ $e$ in the final state. 
Six of these events had an invariant mass of the
two highest-transverse energy leptons ($M_{12}$) greater than 100 GeV, compared to a
SM expectation of $\simeq 0.53$. Such events are predicted in the SM from the Bethe-Heitler
reaction $\gamma \gamma \rightarrow l^+ l^-$, where one of the photons is radiated
from the initial quark and the other from the electron/positron beam. 
An excess could be due to new physics, like for instance doubly-charged Higgs
production.
A coherent combination of H1 and ZEUS data was
performed on the full HERA statistics~\cite{multil}, choosing a common phase space for the
selection. At least
two $e$ candidates in an event had to be central
($20^\circ < \theta^l < 150^\circ$), of which one must have
$P_T^l>10~\mathrm{GeV}$ and the other  $P_T^l>5~\mathrm{GeV}$.
The distribution of the invariant mass $M_{12}$
is shown for the $e^+p$ HERA sample in Fig.~\ref{fig:multileptons}. 
For $M_{12}>100~\mathrm{GeV}$, a total of 4 ($2e$) events
and 4 ($3e$) events were observed in $e^+p$, for an expectation of $1.97 \pm 0.22$ and
$1.10 \pm 0.12$, respectively.
Another interesting distribution is shown in Fig.~\ref{fig:multileptons}, 
the scalar sum of the transverse
momentum $\sum P_T$, where a marginal
excess is seen at values greater than 100 GeV. From Table~\ref{tab:multileptons} it
can be seen that the
excess is in the $e^+p$ data, where 5 events are observed,
while $1.8 \pm 0.2$ are expected. The study of multileptons is being extended
also to muons and taus, the analysis is still in progress.

\section{Summary}
  
The H1 and ZEUS Collaborations are completing the analysis of the cross sections
in NC and CC interactions. The full HERA data set will
be combined and a  more precise determination of the PDFs  
and of the electroweak couplings will derive.

\bigskip
I would like to thank Stefan Schmitt, Matthew Wing and
James Ferrando   for a careful reading of this report
and the organizers for a very enjoyable time in Perugia.

\end{document}